\documentclass[review]{elsarticle}

\usepackage{lineno,hyperref}
\modulolinenumbers[5]
\usepackage{amsmath,amssymb,amsfonts}
\usepackage{algorithmic}
\usepackage{graphicx}
\usepackage{textcomp}
\usepackage{xcolor}
\usepackage{pdfpages}
\usepackage{subfigure}
\usepackage{tabularx}
\usepackage{booktabs,multirow}

\def\BibTeX{{\rm B\kern-.05em{\sc i\kern-.025em b}\kern-.08em
		T\kern-.1667em\lower.7ex\hbox{E}\kern-.125emX}}
\begin{document}

\begin{frontmatter}
\title{MF-JMoDL-Net: A Deep Network for Azimuth Undersampling Pattern Design and Ambiguity Suppression for Sparse SAR Imaging}

\author{
	Yuwei Wu\textsuperscript{a,b,c,d,e},
	Zhe Zhang \textsuperscript {b,c,d,e,*},
	Xiaolan Qiu\textsuperscript{b,c,d,e},
	Yao Zhao\textsuperscript{f},
	Weidong Yu\textsuperscript{a,b,e}
}
\address
{
	\textsuperscript{a }Department of Space Microwave Remote Sensing System,  Beijing, China.\\
	\textsuperscript{b }Aerospace Information Research Institute, Chinese Academy of Sciences, Beijing, China.\\
	\textsuperscript{c }Suzhou Key Laboratory of Microwave Imaging, Processing and Application Technology, Suzhou, Jiangsu, China\\
	\textsuperscript{d }Suzhou Aerospace Information Research Institute, Suzhou, Jiangsu, China\\
	\textsuperscript{e }School of Electronic, Electrical and Communication Engineering, University of Chinese Academy of Sciences, Beijing, China.\\
	\textsuperscript{f }Guangdong University of Technology, Guangzhou, Guangdong, China.\\
	\textsuperscript{* }Corresponding author: Zhe Zhang, Email: zhangzhe01@aircas.ac.cn.\\	
}

%
%
%
%
%

\begin{abstract}
Traditionally, the range swath of a synthetic aperture radar (SAR) system is constrained by its pulse repetition frequency (PRF). Given the system complexity and resource constraints, it is often difficult to achieve high imaging performance and low ambiguity without compromising the swath. In this paper, we propose a joint optimization framework for sparse strip SAR imaging algorithms and azimuth undersampling patterns based on a deep convolutional neural network, combined with matched filter (MF) approximate measurement operators and inverse MF operators, referred to as MF-JMoDL-Net, for sparse SAR imaging methods. Compared with conventional sparse SAR imaging, MF-JMoDL-Net enables us to alleviate the limitations imposed by PRF. In the proposed scheme, joint and continuous optimization of azimuth undersampling patterns and convolutional neural network parameters are implemented to suppress azimuth ambiguity and enhance sparse SAR imaging quality. Experiments and comparisons under various conditions demonstrate the effectiveness and superiority of the proposed framework in imaging results.
\end{abstract}


\end{frontmatter}


\section{Introduction}

%
%
%
%
	Synthetic Aperture Radar (SAR) is a prominent microwave imaging technology, renowned for its all-weather, round-the-clock, and high-resolution capabilities. It has been extensively utilized in remote sensing and monitoring applications \cite{curlander1991SAR}. Both contemporary civilian and military applications demand high spatial resolution. In accordance with the Nyquist sampling theorem, a high pulse repetition frequency (PRF) is necessary to attain a high azimuth resolution. Regrettably, an elevated PRF escalates the volume of echo data, potentially exceeding the on-board storage capacity and downlink bandwidth of a satellite. Consequently, an efficient azimuth undersampling scheme that minimizes azimuth ambiguity is of paramount importance for SAR systems.
	
	To mitigate azimuth ambiguity, various signal processing methods have been proposed. The in-phase cancellation technique introduced in \cite{moreira_suppressing_1993} alters the phase of the ambiguity signal and reduces azimuth ambiguity by constructing a correction filter. The restoration method \cite{Chen2014} employs a defined metric, SAR system parameters, and the local mean energy of SAR images to identify ambiguity pixels, followed by selecting an ambiguity region restoration mechanism based on the size of ambiguity regions. The inpainting method proposed in \cite{long_azimuth_2020} combines Wiener filtering and local azimuth ambiguity ratio (AASR) estimation to suppress N-order azimuth ambiguity. While this method can eliminate N-order azimuth ambiguity, it is unable to flexibly recover the actual targets affected by ambiguity. Nonetheless, all these methods focus on suppressing azimuth ambiguity in the context of uniform sampling rather than nonuniform sampling.
	
	In contrast to uniform sampling, stochastic sampling has been introduced and investigated by several researchers \cite{zhang2005stochastic,Shapiro1960,Masry1978}. Subsequently, stochastic sampling patterns were adopted in computer graphics for antialiasing, such as jittered sampling and Poisson disk sampling \cite{Dippe1985,Cook1986}. Villano et al. were the first to introduce staggered sampling \cite{Villano2012} into SAR imaging, termed staggered-SAR, where the PRF was continuously varied \cite{Villano2014a,Villano2014b,Villano2017}, and ambiguity in the image manifested as noise-like disturbances rather than localized artifacts \cite{Villano2015}. Regrettably, this processing method necessitates the system to operate at a high azimuth oversampling rate, resulting in significantly increased data rates for SAR hardware systems.
	
	Compressive sensing (CS) theory \cite{Candes2006, Donoho2006} has been introduced into radar imaging \cite{Baraniuk2007, rilling2009compressed}, enabling SAR imaging at low sampling rates (CS-SAR). Zhang et al. \cite{Zhang2012}, He et al. \cite{He2012}, and Cetin et al. \cite{Cetin2014} proposed sparse SAR imaging, which reconstructs the scene of interest by solving a regularization problem using CS recovery algorithms, such as Orthogonal Matching Pursuit (OMP) \cite{Tropp2007}, Compressive Sampling Matching Pursuit (CoSaMP) \cite{Needell2009a, Needell2010}, Fast Iterative Shrinkage-Thresholding Algorithm (FISTA) \cite{Beck2009}, and Iterative Shrinkage-Thresholding Algorithm (ISTA) \cite{Daubechies2004}. These algorithms enable direct processing of downsampling raw data. However, in the context of satellite SAR imaging, handling a large measurement matrix while processing raw data is challenging. To address this limitation, an efficient regularization algorithm for SAR sparse imaging is proposed, which firstly suggested that the matched filtering (MF) algorithm could be considered an imaging operator, thus forming a new CS-SAR imaging method for high-quality, high-resolution imaging under sub-Nyquist rate sampling \cite{Fang2014, Jiang2014}. With the support of sparse SAR imaging technology, the possibility of azimuth undersampling has garnered researchers' attention.
	Sun et al. \cite{Sun2012} proposed a CS-SAR imaging approach based on azimuth Poisson disk-like nonuniform sampling, which distinguishes itself from staggered-SAR by adopting a sampling pattern based on Poisson downsampling rate instead of linear oversampling rate. This method significantly reduces the computational complexity of CS reconstruction from $\mathcal{O}(2MN)$ to $\mathcal{O}(4N\log N)$, but it only corrects the range walk during the range cell migration correction (RCMC) process without addressing the range curvature. To resolve this issue, \cite{Yang2019, Yang2019a} proposed an algorithm that combines the CSA operator and ISTA to correct both the range walk and the range curvature under Poisson disk nonuniform downsampling conditions. Although sparse SAR imaging offers the aforementioned advantages, it also introduces nonuniform sampling, echo loss, and non-ideal AAP issues, leading to considerable aliasing and ambiguity particularly in azimuth direction.
	
	Deep learning has achieved remarkable performance in signal and image processing, with the deep unrolling algorithm demonstrating the potential for network-based SAR imaging reconstruction \cite{Monga2021}. Mason et al. \cite{Mason2017} introduced a recurrent autoencoder network architecture based on the ISTA that incorporated SAR modeling, exhibiting faster convergence and reduced reconstruction error compared to standard iterative analytic methods. Pu et al. \cite{Pu2021a} proposed an autoencoder-structured deep SAR imaging algorithm, where motion compensation is considered to mitigate the impact of motion errors on imaging results. In \cite{Zhao2021}, an end-to-end SAR deep learning imaging method based on a sparse optimization iterative algorithm was suggested to enhance the universality and generalization ability of the imaging method for SAR echo data. Li et al. \cite{Li2022} developed a target-oriented SAR imaging model, where generalized regularization was employed to characterize target features, contributing to an improved signal-to-clutter ratio (SCR) in the reconstructed image. It is evident that existing deep network-based SAR imaging methods primarily concentrate on enhancing the quality of SAR images. However, no research combines the deep network and azimuth sampling pattern in sparse SAR imaging. Current deep network-based sparse SAR imaging works typically directly adopt canonical uniform or random undersampling schemes, but these undersampling patterns are suboptimal in most cases. 
	In the biomedical imaging field, Aggarwal et al. \cite{Aggarwal2020} proposed a joint model-based deep learning approach for optimizing downsampling and reconstruction (J-MoDL) specifically for magnetic resonance imaging (MRI), based on an image reconstruction framework with a CNN-based regularization prior \cite{Aggarwal2019}. This method significantly enhances imaging quality and the performance of most deep learning reconstruction algorithms.
	
	Inspired by J-MoDL, this paper proposes a joint optimization framework for sparse strip SAR imaging algorithms and azimuth undersampling patterns based on a deep convolutional neural network combined with an MF approximate measurement operator. The main contributions of this article are as follows:
	
	1) \emph{A learnable undersampling pattern framework is proposed}: This is the first time a deep learning model has been introduced into the optimization of SAR azimuth sampling patterns. In this paper, we design a data-driven azimuth sampling model that extracts features using the U-Net network \cite{Ronneberger2015}. Leveraging the exceptional information extraction capability of deep neural networks, effective information is separated from clutter information, which contributes to obtaining more information of interest while using as few pulses as possible. The proposed sampling method outperforms existing azimuth pulse repetition interval (PRI) designs in terms of imaging performance.
	
	2) \emph{A deep network imaging approach based on nonuniform MF operator is presented}: The MF-JMoDL-Net utilizes an approximate measurement operator to significantly reduce storage and computation costs. We have originally incorporated the nonuniform fast Fourier transform (NUFFT) MF operator into the network-based SAR imaging framework, making the proposed method efficient and feasible. Furthermore, the backpropagation based on conjugate gradients can enforce data consistency, unlike conventional steepest descent updates. Its ability to work with smaller CNN modules allows it to learn from smaller datasets, making it well-suited to the issue of small SAR imaging training datasets in deep learning applications.
	
	3) \emph{Experiments are conducted under various conditions}: To verify the effectiveness of the proposed method, we make comparisons among the proposed algorithm, Poisson-disk, and staggered sampling patterns, and perform experiments on conventional azimuth ambiguity suppression. All the results demonstrate the feasibility and reliability of the proposed method. Additionally, we investigate the preliminary analysis of the obtained sampling patterns for different scenes, and explore the imaging performance at 50\%, 25\%, and 12.5\% undersampling rates, aiming to discuss how this novel SAR imaging method performs on ambiguity suppression as the sampling rate decreases.
	
	The rest of this paper is organized as follows: Section \ref{sec:2} presents the SAR signal model and the proposed imaging model. Section \ref{sec:3} describes the solution of the imaging model and provides details on the MF-JMoDL-Net. In Section \ref{sec:4}, experimental results and related analysis are presented. Finally, conclusions are drawn in Section \ref{sec:5}.

\begin{figure}[t]
	\centering
	\includegraphics[width=0.7\linewidth]{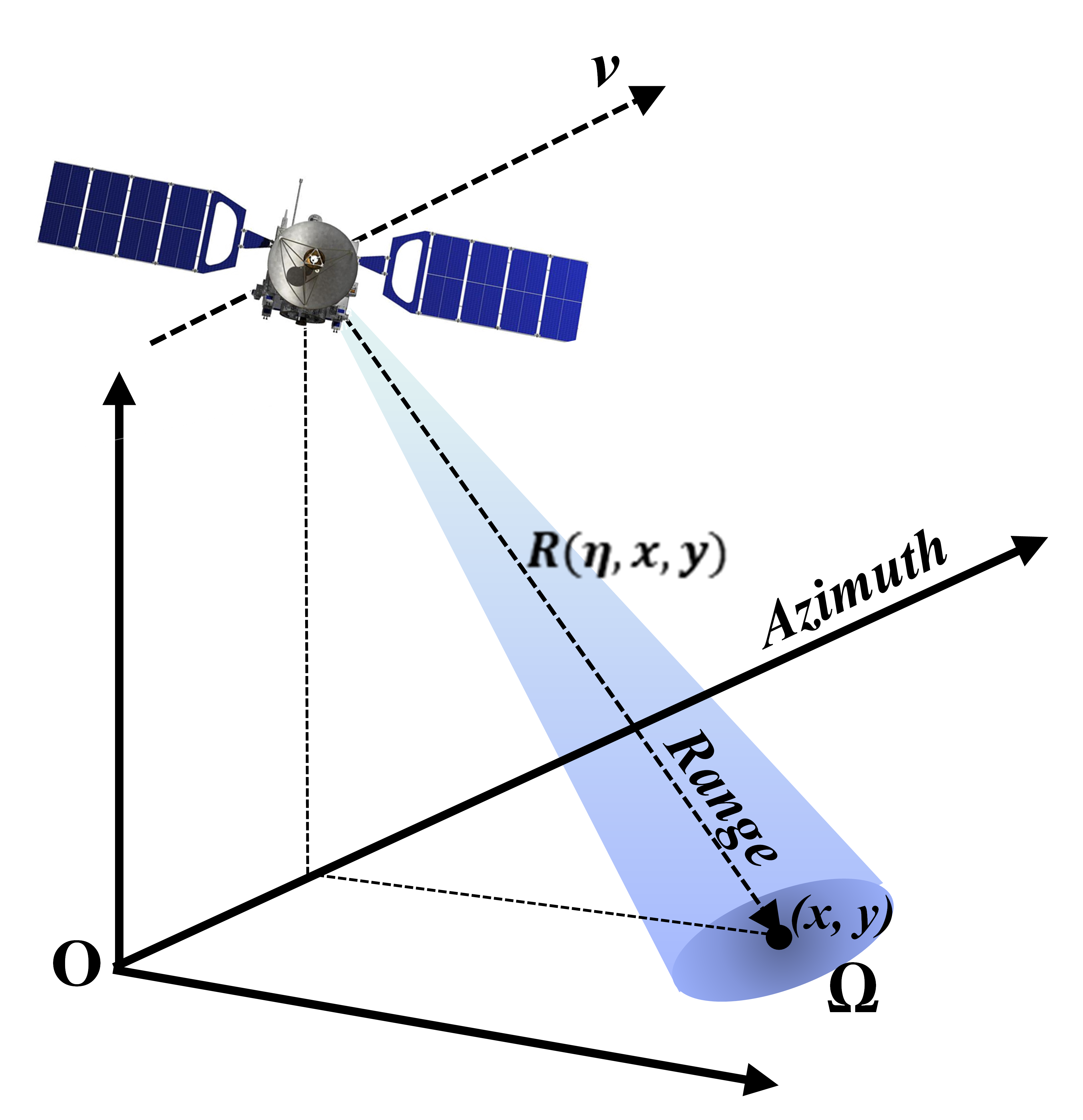}
	\caption{Geometric configuration of side view stripmap mode SAR.}
	\vspace{-2mm}
	\label{fig:Fig1}
\end{figure}

\section{Problem Setting and Approaches}\label{sec:2}
In this section, we first provide the stripmap SAR signal model. Then, in Subsection \ref{sec:2.2}, the joint optimized sampling scheme and reconstruction framework are presented to clarify the intention of the proposed SAR imaging model. Finally, in Subsection \ref{sec:2.3}, the nonuniform MF-based approximate measurement model is introduced for the completeness of this article.
\subsection{Stripmap SAR Signal Model}\label{sec:2.1}

We assume that the SAR system operates in single-channel strip mode. The geometric configuration of a typical side-looking SAR is shown in Fig.\ref{fig:Fig1} Suppose that the platform carrying the imaging radar moves in a straight line relative to the observation scene at a constant speed $v$. For stripmap SAR, the direction of the antenna beam remains fixed throughout the entire movement. Consequently, the distance between the phase center of the antenna and the target point with coordinates $(x,y)$ in the observation scene is
\begin{equation}\label{1}
	R\left( {\eta,x,y} \right) = \sqrt{(x - v\eta)^{2} + R_{0}^{2}}
\end{equation}
where $\tau$ and $\eta$ are the fast time and the slow time, respectively, $x$ and $y$ are the spatial coordinates of the target point along the azimuth and range directions, and $R_{0}$ is the closest range between target and platform.

The signal transmitted by the antenna to the observation scene is a linear frequency modulated signal (LFM), therefore, the echo received by the imaging radar of the target point $(x,y)$ in the observation scene is
\begin{equation}\label{2}
	\begin{split}	
		s\left( {\tau,\eta,x,y} \right) = A_{0}\sigma\left( {x,y} \right) \cdot \omega_{r}\left( {\tau - \frac{2R\left( {\eta,x,y} \right)}{c}} \right)  \\
		\cdot exp\left\{ {j\pi K_{r}\left( {\tau - \frac{2R\left( {\eta,x,y} \right)}{c}} \right)^{2}} \right\}\cdot \omega_{a}\left( {\eta - \eta_{c}} \right) \\
		\cdot {\mathit{\exp}\left\{ {- \frac{j4\pi f_{0}R\left( {\eta,x,y} \right)}{c}} \right\}}
	\end{split}
\end{equation}
where $A_0$ is the observation gain determined by the antenna; $\sigma\left( {x,y} \right)$ is the backscattering coefficient of the target point $(x,y)$; $\omega_{a}( \cdot )$ and $\omega_{r}( \cdot )$ are the azimuth and range signal envelope, respectively; $f_0$ is the carrier frequency; $c$ is the speed of light. The SAR echo reflected from the scatterers  generated by the antenna beam coverage area $\Omega$ can be expressed as 
\begin{equation}\label{3}
	\begin{split}
		s\left( {\tau,\eta} \right) = {\iint_{x,y \in \Omega}{s\left( {\tau,\eta,x,y} \right)dxdy}} + N\left( {\tau,\eta} \right)
	\end{split}
\end{equation}
where $N(\tau,\eta)$ is the additive measurement noise. By discretizing the observation scene $\Omega$ into $M_{r} \times M_{a}$ grids and sampling the echo signal into a matrix, the discrete echo can be expressed as
\begin{equation}\label{4}
	\begin{split}
		s\left( {\tau_{n_{r}},\eta_{n_{a}}} \right) =  {\sum\limits_{m_{r}}^{M_{r}}{\sum\limits_{m_{a}}^{M_{a}} h\left\lbrack \tau_{n_{r}},\eta_{n_{a}},x_{m_{r}},y_{m_{a}} \right\rbrack}} \\
		\cdot{\sigma\left( {x_{m_{r}},y_{m_{a}}} \right)} + N\left( \tau_{n_{r}},\eta_{n_{a}} \right)
	\end{split}
\end{equation}
where $s\left( {\tau_{n_{r}},\eta_{n_{a}}} \right)$ is the $n_{r}$-th range sample at the $n_{a}$-th pulse of raw signal $s\left( {\tau,\eta} \right)$ , $M_r$ and $M_a$ are the numbers of the discrete cells in the range and azimuth directions, respectively. The $h\left\lbrack \tau_{n_{r}},\eta_{n_{a}},x_{m_{r}},y_{m_{a}} \right\rbrack$ is defined by
\begin{equation}\label{5}
	\begin{split}
		h\left\lbrack {\tau_{n_{r}},\eta_{n_{a}},x_{m_{r}},y_{m_{a}}} \right\rbrack =
		\omega_{r}\left( {\tau_{n_{r}} - \frac{2R\left( {\eta_{n_{a}},x_{m_{r}},y_{m_{a}}} \right)}{c}} \right) \\
		\cdot exp\left\{ {j\pi K_{r}\left( {\tau_{n_{r}} - \frac{2R\left( {\eta_{n_{a}},x_{m_{r}},y_{m_{a}}} \right)}{c}} \right)^{2}} \right\}\\
		\cdot\omega_{a}\left( {\eta_{n_{a}} - \eta_{c}} \right)
		\cdot {\mathit{\exp}\left\{ {- \frac{j4\pi f_{0}R\left( {\eta_{n_{a}},x_{m_{r}},y_{m_{a}}} \right)}{c}} \right\}}
	\end{split}
\end{equation}

Write the discretized data $s\left( {\tau_{n_{r}},\eta_{n_{a}}} \right)$ and the reflectivity map $\mathbf{\sigma}\left( {x_{m_{r}},y_{m_{a}}} \right)$ as two matrices: $\mathbf{S} \in \mathbb{C}^{{M \times N}_{r}}$ and $\mathbf{\sigma} \in \mathbb{C}^{{M_{a} \times M}_{r}}$, respectively, where $M$ is the number of transmitted pulses in the azimuth direction; $N_r$ is the number of samples in the range direction. Then, (\ref{4}) can be rewritten as
\begin{equation}\label{6}
	\begin{split}
		\mathbf{S} = \mathbf{H} \mathbf{\sigma} + \mathbf{n}
	\end{split}
\end{equation}
where $\mathbf{n}$ is the vector of additive noise, and $\mathbf{H}$ is defined by
\begin{equation}\label{7}
	\mathbf{H} = 
	\begin{bmatrix}
		{a\left\lbrack \tau_{1},\eta_{1},x_{1},y_{1} \right\rbrack} & \cdots & {a\left\lbrack \tau_{1},\eta_{1},x_{M_{a}},y_{M_{r}} \right\rbrack} \\
		{a\left\lbrack \tau_{1},\eta_{2},x_{1},y_{1} \right\rbrack} & \cdots & {a\left\lbrack \tau_{1},\eta_{2},x_{M_{a}},y_{M_{r}} \right\rbrack} \\
		\vdots & \ddots & \vdots \\
		{a\left\lbrack \tau_{1},\eta_{M},x_{1},y_{1} \right\rbrack} & \cdots & {a\left\lbrack \tau_{1},\eta_{M},x_{M_{a}},y_{M_{r}} \right\rbrack} \\
		\vdots & \ddots & \vdots \\
		{a\left\lbrack \tau_{N_{r}},\eta_{M},x_{1},y_{1} \right\rbrack} & \cdots & {a\left\lbrack \tau_{N_{r}},\eta_{M},x_{M_{a}},y_{M_{r}} \right\rbrack} \\
	\end{bmatrix}.
\end{equation}

Conventionally, the intention of CS-SAR imaging is to reconstruct $\mathbf{\sigma}$ from the measurements $\mathbf{S}$.

\subsection{MF-Based Image Formation and Proposed Forward Model}\label{sec:2.2}
In this subsection, we will introduce and discuss the typical MF-Based CS-SAR imaging model to explain the original intention of the proposed method. Then, we discuss the forward model with the sampling pattern, which is the central assumption of the proposed MF-JMoDL-Based SAR imaging model.

1) \emph{Typical MF-Based CS-SAR Imaging Model}: in the typical CS-SAR imaging model, the measurement vector $\mathbf{H}$ is generally resampled by an appropriate sampling matrix $\Theta$. When the observation scene is sparse and the observation matrix $\mathbf{A} = \Theta\mathbf{H}$ satisfies the RIP condition \cite{Candes2006}, $\mathbf{\sigma}$ can be exactly recovered from $\mathbf{S}$ with the $L_{q}(0 < q < 1)$ optimization:
\begin{equation}\label{8}
	\begin{split}
		{\min\limits_{\mathbf{\sigma}}\left\| \mathbf{\sigma} \right\|_{q}}~~~~~~~s.t.~~\mathbf{S} = \mathbf{A}\mathbf{\sigma}
	\end{split}
\end{equation}

To solve this problem, an equivalent regularization scheme with the following optimization problem can be used:
\begin{equation}\label{9}
	\begin{split}
		\underset{\mathbf{\sigma}}{min~}\left\{ {\left\| {\mathbf{S} - \mathbf{A}\mathbf{\sigma}} \right\|_{2}^{2} + \lambda\left\| \mathbf{\sigma} \right\|_{q}^{q}} \right\}
	\end{split}
\end{equation}
where $\lambda$ is a regularization parameter. Generally speaking, the solution methods for (\ref{8}) and (\ref{9}) can be divided into two categories: greedy methods and iterative methods. However, regardless of the chosen method, the computational and storage costs are enormous due to the size of the measurement matrix $\mathbf{H}$, which significantly reduces computational efficiency.
Hence, the corresponding reconstruction algorithms are considerably more time-consuming than traditional matched filter MF-based focusing methods. 

To address this issue, the MF-based sparse SAR imaging model \cite{Jiang2014, Fang2014} offers a general principle on how the observation can be remodeled and approximated by any high-precision imaging procedure:
\begin{equation}\label{10}
	\begin{split}
		\mathbf{G} = \mathbf{M}^{- 1} \approx \mathbf{H}
	\end{split}
\end{equation}
where $\mathbf{M}$ is the traditional MF imaging procedure that can be calculated through decoupling it into a series of 1-D operators in the frequency domain, like chirp scaling algorithm (CSA) and range doppler algorithm (RDA), $\mathbf{G}$ is any generalized right inverse of $\mathbf{M}$.

2) \emph{Forward Model with Sampling Pattern}: To optimize the azimuth sampling scheme of MF-based SAR imaging, we employ the model-based imaging algorithms \cite{Ongie2015}\cite{Ongie2018}. The model-based imaging schemes use a continuous function of $\mathbf{A}$, denoted by the operator $H_{\Theta_{a}}^{*}( \cdot )$, which maps $\mathbf{\sigma}$ to $\mathbf{S}$, and $\Theta_{a}$ represents the azimuth sampling pattern. Then, we construct the forward model as follows:
\begin{equation}\label{11}
	\begin{split}
		\mathbf{S} = H_{\Theta_{a}}^{*}\left( \mathbf{\sigma} \right)
	\end{split}
\end{equation}

Due to the recovery of $\mathbf{\sigma}$ from $\mathbf{S}$ is ill-posed, model-based imaging algorithms pose the recovery as a regularized optimization scheme:
\begin{equation}\label{12}
	\begin{split}
		\mathbf{\sigma} = {\arg{\min\limits_{\mathbf{\sigma}}{\underset{\rm data~consistency}{\underbrace{\left\| {\mathbf{S} - H_{\Theta_{a}}^{*}\left( \mathbf{\sigma} \right)} \right\|_{2}^{2}}} + \lambda\underset{\rm regularization}{\underbrace{N\left( \mathbf{\sigma} \right)}}}}}
	\end{split}
\end{equation}
here $N\left( \mathbf{\sigma} \right)$ is a regularization penalty. When $\mathbf{\sigma}$ is an artifact-free image, $N\left( \mathbf{\sigma} \right)$ is a small scalar, otherwise, it is a high scalar for noisy images. Commonly-used regularizers in imaging reconstruction include total variation regularization \cite{Ma2008}, transform domain sparsity \cite{Figueiredo2003}, and structured low-rank methods \cite{Jacob2020}. These regularizers aim to enforce specific desirable properties on the reconstructed image, such as smoothness, sparsity in certain domains, or low-rank structures, ultimately leading to improved image quality and reduced artifacts.

\subsection{Nonuniform MF Operator}\label{sec:2.3}
\begin{figure}[!t]
	\centering
	\subfigure[CSA operator]{
		\includegraphics[width=0.7\linewidth]{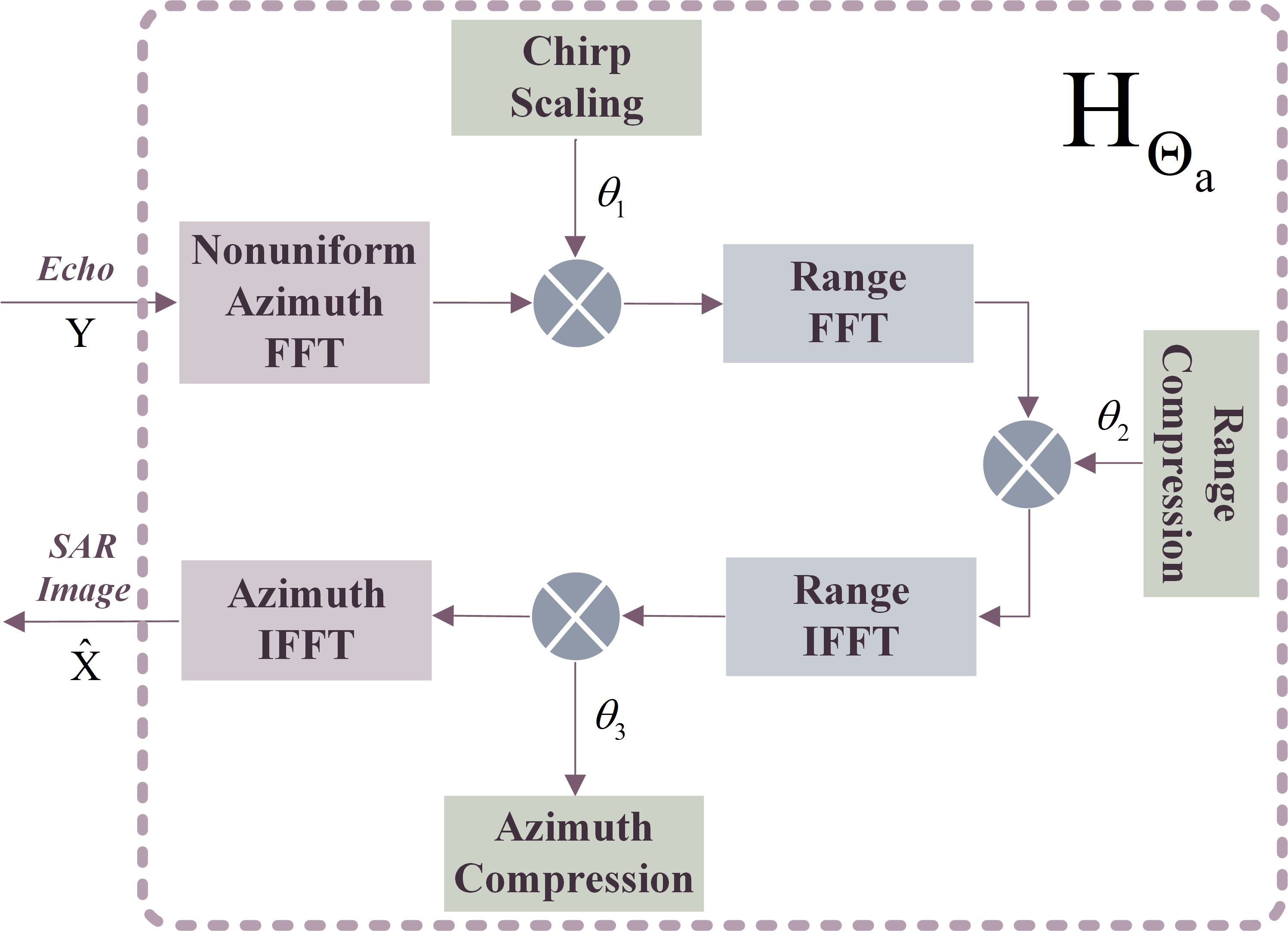}
	}\label{fig:Fig2(a)}
	\subfigure[Inverse CSA operator]{
		\includegraphics[width=0.7\linewidth]{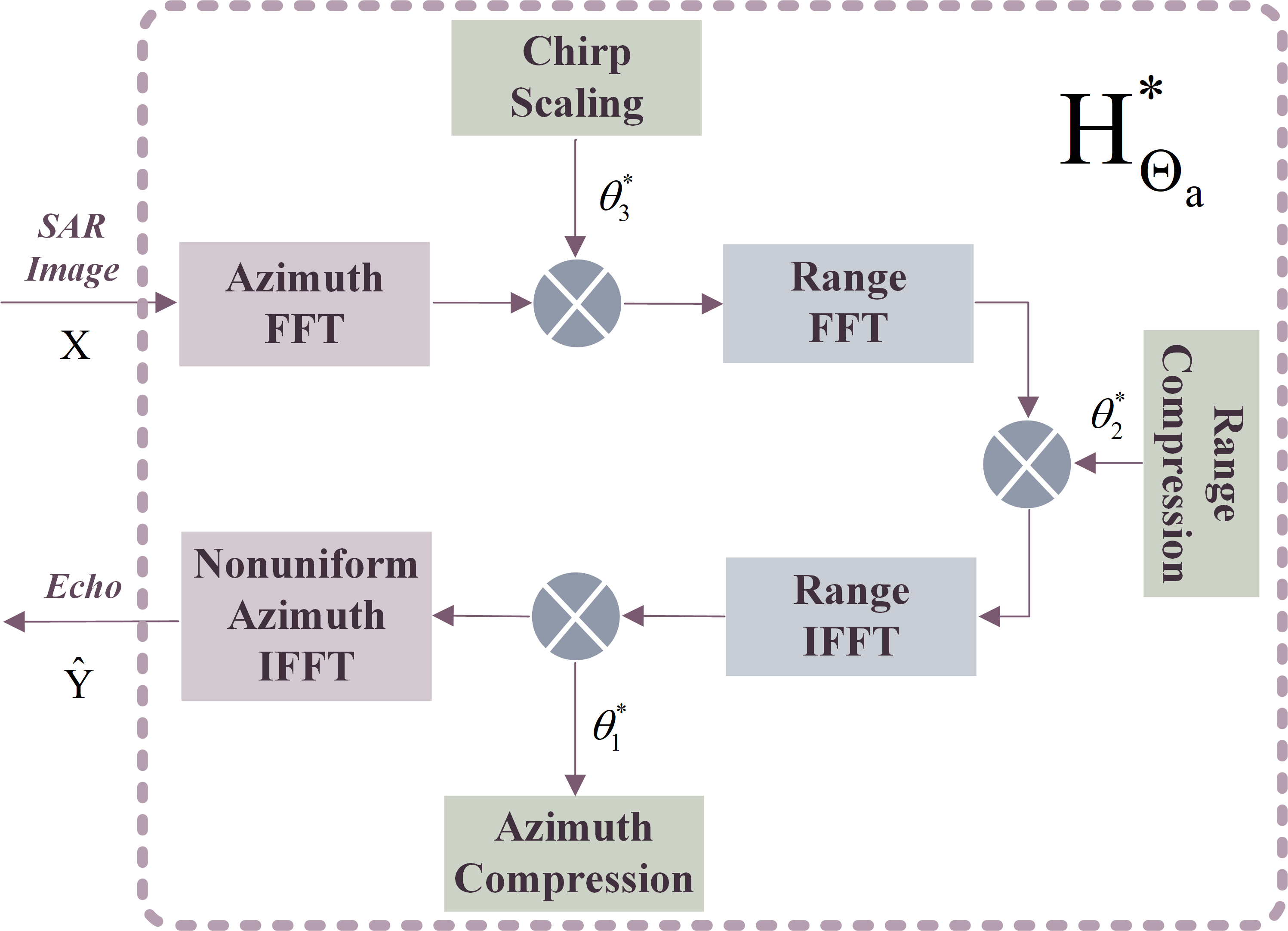}
	}\label{fig:Fig2(b)}
	\caption{Procedure of CSA operator and inverse CSA operator.}
	\label{fig:Fig2}
\end{figure}

In order to address the challenges posed by the proposed model in \ref{sec:2.2}, two main problems must be overcome.

Firstly, the large-scale exact measurement matrix $\mathbf{A}$ presents significant computational and storage requirements, which would reduce the efficiency of the proposed SAR imaging model. To tackle this issue, the MF-based SAR imaging method is employed to construct an approximate measurement operator by decoupling azimuth and range processing. Following a similar approach to the work in \cite{Jiang2014}, the chirp scaling algorithm (CSA) is used to build the approximate measurement operator.

Secondly, the sampling pattern proposed in this paper leads to nonuniform samples in the azimuth direction. Consequently, a nonuniform MF operator $H_{\Theta_{a}}( \cdot )$ is needed, which is different from the regular MF operator. In a similar manner to the work in \cite{Yang2019a}, the chirp scaling algorithm (CSA) is adopted to construct the approximate measurement operator. The process for constructing the approximate measurement operator is illustrated in Fig.\ref{fig:Fig2}(a) and Fig.\ref{fig:Fig2}(b).

By addressing these two challenges, the proposed SAR imaging model can be solved efficiently while maintaining the benefits of the optimized azimuth sampling pattern.

The formulation of CSA can be expressed as follows:
\begin{equation} \label{13}
	\begin{split}
		\hat{\mathbf{\sigma} } = \mathbf{I}{\mathbf{F}_{\eta} }\left( {\mathbf{I}{\mathbf{F}_{\tau} }\left( {{\mathbf{F}_{\tau} }\left( {\mathbf{N}{\mathbf{F}_{{{\eta} _{{{\rm{\Theta }}_a}}}}}\left( \mathbf{S} \right) \odot {{\theta} _1}} \right) \odot {{\theta} _2}} \right) \odot {{\theta} _3}} \right)
	\end{split}
\end{equation}
In the given context, $\hat{\mathbf{\sigma}}$ represents the reconstructed SAR image matrix, while $\mathbf{F}{\tau}$ and $\mathbf{F}{\eta}$ denote the fast Fourier transform (FFT) along the range direction and azimuth direction, respectively. $\mathbf{IF}{\tau}$ is the inverse fast Fourier transform (IFFT) along the range direction. The phase terms for RCMC, range compression, azimuth-range decoupling, and azimuth compression are represented by filters $\theta{1}$, $\theta_{2}$, and $\theta_{3}$, respectively. The expressions for these phase terms can be found in \cite{Yang2019a}, and as such, they are not repeated here.

The equivalent CSA operator is denoted as $H_{\Theta_{a}}( \cdot )$. Therefore, equation (\ref{13}) can be expressed as $\hat{\mathbf{\sigma}} = H_{\Theta_{a}}\left( \mathbf{S} \right)$. This formulation allows for efficient processing of the SAR imaging problem while taking into account the optimized azimuth sampling pattern.

Now we defined the nonuniform Fourier transform matrix $\mathbf{NF}_{\eta_{\Theta_{a}}}$ in (\ref{10}), which is used to transform the raw data into the range–doppler domain as
\begin{equation} \label{14}
	\begin{split}
		{\mathbf{N}\mathbf{F}}_{\eta_{\Theta_{a}}} = \left\lbrack {\mathbf{\alpha}_{1},\mathbf{\alpha}_{2},\ldots,\mathbf{\alpha}_{\mathbf{M}}} \right\rbrack
	\end{split}
\end{equation}
\begin{figure*}[t]
	\centering
	\includegraphics[width=0.95\linewidth]{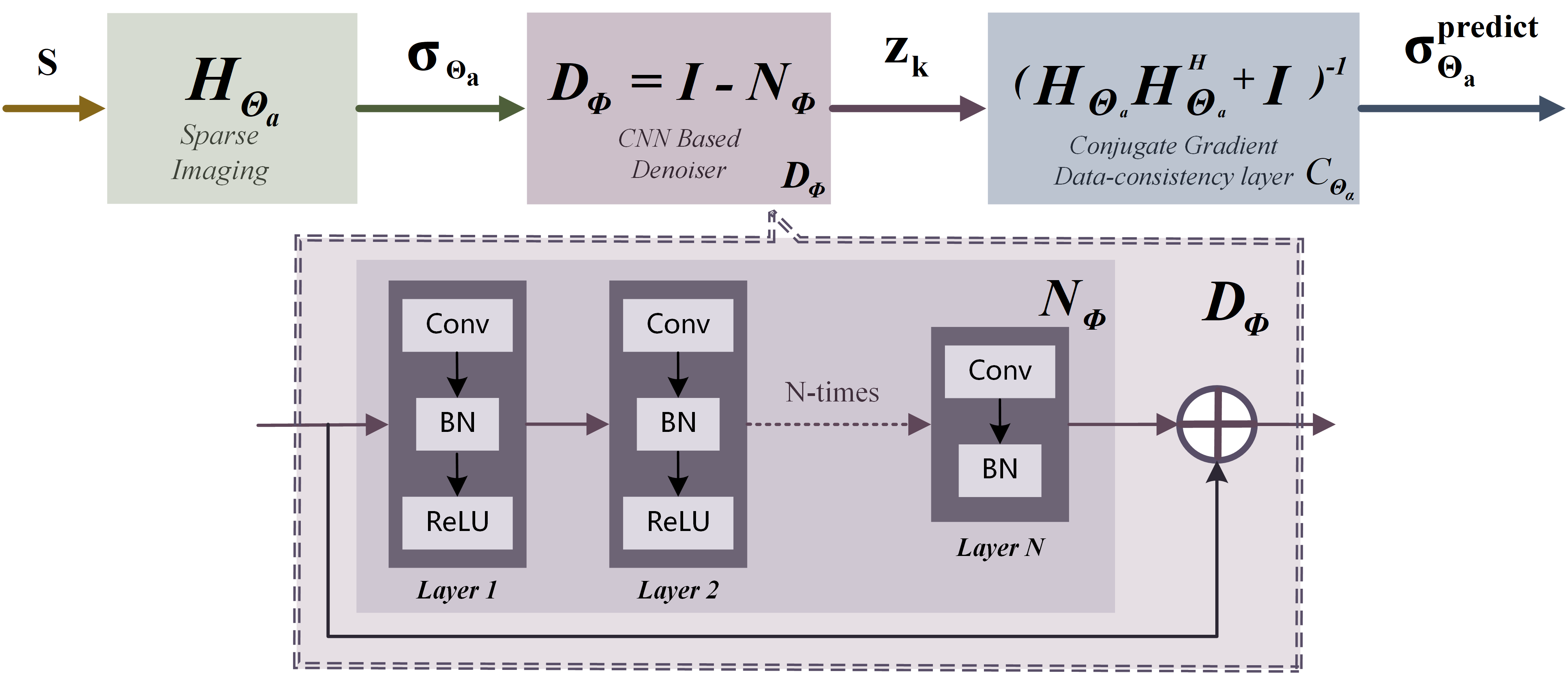}
	\caption{The recursive framework that alternates between CNN based denoiser block $D_{\Phi}$ in (\ref{25}) and the data-consisitency (DC) layer in (\ref{26}).}
	\vspace{-2mm}
	\label{fig:Fig3}
\end{figure*}
it consists of $M$ column vectors defined by
\begin{equation} \label{15}
	\begin{split}
		{\alpha _m} = {\left[ {exp\left( { - j2\pi {\eta _m}f_\eta ^{\left( 1 \right)}} \right), \ldots ,exp\left( { - j2\pi {\eta _m}f_\eta ^{\left( {{N_a}} \right)}} \right)} \right]^T}
	\end{split}
\end{equation}

In equation (\ref{15}), $\eta_{m}$ represents the $m$-th sample in the azimuth direction, with $m = 1, 2, \dots, M$, where $M$ is the number of samples in azimuth. The $n$-th sample in Doppler frequency is given by $f_{\eta}^{(n)} = (n - 1)PRF/N_{a}$, where $n = 1, 2, \dots, N_{a}$, and $N_{a}$ is the number of uniform frequency grids in the Doppler domain. $PRF$ denotes the azimuth pulse repetition frequency.

By applying the CSA operator to the raw data $\mathbf{S}$, we can obtain the reconstructed backscattering coefficients of the scatterers in the observed scene. These coefficients represent the true backscattering coefficients we desire, which are approximately equal to $\mathbf{\sigma}$. The expression is as follows:
\begin{equation} \label{16}
	\begin{split}
		H_{\Theta_{a}}\left( \mathbf{S} \right) = \hat{\mathbf{\sigma}} \approx \mathbf{\sigma}
	\end{split}
\end{equation}

From (\ref{13}), it is evident that the CSA operator is invertible, as mentioned in \cite{Yang2019a}. By applying the inverse of the CSA operator to the true backscattering coefficients of the scene, we can obtain an approximation of the raw data $\mathbf{S}$. The inverse CSA operator can be explicitly expressed as follows:
\begin{equation} \label{17}
	\begin{split}
		\hat{\mathbf{S}} = {\mathbf{N}\mathbf{I}\mathbf{F}}_{\eta_{\Theta_{a}}}\left( \mathbf{I}\mathbf{F}_{\tau}\left( \mathbf{F}_{\tau}\left( \mathbf{F}_{\eta}\left( \mathbf{\sigma} \right) \odot \theta_{3}^{\mathbf{*}} \right) \odot \theta_{2}^{\mathbf{*}} \right) \odot \theta_{1}^{\mathbf{*}} \right)
	\end{split}
\end{equation}
where $\mathbf{F}_{\eta}$ is the azimuth FFT operator. $\hat{\mathbf{S}}$ is the approximated echo data matrix generated by $H_{\Theta_{a}}( \cdot )$, then the (\ref{17}) can be expressed as $\hat{\mathbf{S}} = H_{\Theta_{a}}\left( \mathbf{\sigma} \right)$. the nonuniform inverse Fourier transform ${\mathbf{N}\mathbf{I}\mathbf{F}}_{\eta_{\Theta_{a}}}$ composed of $\mathbf{N}_{\mathbf{a}}$ column vectors defined by
\begin{equation} \label{18}
	\begin{split}
		{\mathbf{N}\mathbf{I}\mathbf{F}}_{\eta_{\Theta_{a}}} = \left\lbrack {\mathbf{\beta}_{1},\mathbf{\beta}_{2},\ldots,\mathbf{\beta}_{\mathbf{N}_{\mathbf{a}}}\mathbf{~}} \right\rbrack
	\end{split}
\end{equation}
which is consist of $\mathbf{N}_{\mathbf{a}}$ column vectors defined by
\begin{equation} \label{19}
	\begin{split}
		\mathbf{\beta}_{\mathbf{n}} = \frac{1}{N_{a}}\left\lbrack exp\left( {j2\pi\eta_{1}f_{\eta}^{(n)}} \right),\ldots,exp\left( {j2\pi\eta_{M}f_{\eta}^{(n)}} \right) \right\rbrack^{T}
	\end{split}
\end{equation}
where $\eta_{1},\ldots,\eta_{M}$ are obtained according to proposed method.

\section{Proposed Method}\label{sec:3}

In this section, we aim to solve the proposed model-based SAR imaging framework from Subection \ref{sec:2.2}. The process consists of four main parts:

1) \emph{Deriving a deep learning framework (Subsection \ref{sec:3.1})}: The deep learning framework is derived to facilitate the solution of the model-based SAR imaging framework. This will enable the efficient processing and optimization of SAR imaging using advanced deep learning techniques.

2) \emph{Joint optimization of sampling pattern (Subsection \ref{sec:3.2})}: We present the joint optimization of the azimuth sampling scheme in SAR imaging. This optimization allows the system to learn an effective sampling pattern for improved SAR imaging performance.

3) \emph{Unfolding the iterative solution into a deep network-based solution, MF-JMoDL-Net (Subsection \ref{sec:3.3})}: The iterative solution is unfolded into a deep network-based solution called MF-JMoDL-Net. This network will be capable of handling the SAR imaging problem with improved efficiency and performance compared to traditional iterative methods.

4) \emph{Loss function, backpropagation, and gradient calculation analysis (Subsection \ref{sec:3.4})}: In this part, we provide details on the loss function used for optimization, the backpropagation process for updating the network weights, and the gradient calculation. This information is crucial for the successful training and implementation of the MF-JMoDL-Net.

By addressing these four components, we will develop a comprehensive solution for the proposed model-based SAR imaging framework, leveraging the power of deep learning to improve the overall performance of SAR imaging systems.

\subsection{JModel-Based Deep Learning Image Reconstruction}\label{sec:3.1}
Consider the model proposed in \ref{sec:2.2}, we define the regularizer as $N\left( \mathbf{\sigma} \right) = \lambda\left\| {\mathbf{T}\mathbf{\sigma}} \right\|_{\mathcal{L}_{1}}$ in transform domain sparsity, with $\Phi = \left\{ \mathbf{T},\lambda \right\}$ denoting the parameters of the transform and regularization. To denote the dependence on the regularization parameters as well as the sampling pattern of (\ref{12}), we give the optimization problem of the form
\begin{equation} \label{20}
	\begin{split}
		\hat{\mathbf{\sigma}}\left\{ {\Phi,\Theta}_{a} \right\} = {\arg{\underset{\mathbf{\sigma}}{min~}{\left\| {\mathbf{S} - H_{\Theta_{a}}^{*}\left( \mathbf{\sigma} \right)} \right\|_{2}^{2} +}}}N_{\Phi}\left( \mathbf{\sigma} \right)
	\end{split}
\end{equation}

To address (\ref{13}), we treat the image reconstruction as a regularization optimization problem and employ a deep learning approach rather than traditional fixed prior regularization to learn parameters from data samples with optimized azimuth sampling patterns, which are commonly referred to as data-driven methods. Classical data-driven image reconstruction techniques include direct inversion schemes \cite{Chen2017a}\cite{Han2020a} and the model-based deep learning framework (MoDL) \cite{Aggarwal2019}. Owing to the substantial reduction in the number of network parameters in MoDL, we focus on this deep learning framework, formulating its image recovery as follows:
\begin{equation} \label{21}
	\begin{split}
		\hat{\mathbf{\sigma}}\left\{ {\Theta_{a},\Phi} \right\} = {\arg{\underset{\sigma}{min~}{\left\| {\mathbf{S} - H_{\Theta_{a}}^{*}\left( \mathbf{\sigma} \right)} \right\|_{2}^{2} + \left\| {\mathbf{\sigma} - D_{\Phi}\left( \mathbf{\sigma} \right)} \right\|_{2}^{2}}}}
	\end{split}
\end{equation}
where $D_{\Phi}\left( \mathbf{\sigma} \right)$ is the ``denoised'' version of $\mathbf{\sigma}$. To extract the ambiguity and noise in $\mathbf{\sigma}$, we let $D_{\Phi}$ be a residual learning-based CNN.
Similar to \cite{Aggarwal2019}, setting $\mathbf{\sigma}_{n} + \mathrm{\Delta}\mathbf{\sigma} = \mathbf{\sigma}$, the non-linear mapping $D_{\Phi}\left( {\mathbf{\sigma}_{n} + \mathrm{\Delta}\mathbf{\sigma}} \right)$ can be approximated using Taylor series around the $n^{th}$ iterate as
\begin{equation} \label{22}
	\begin{split}
		D_{\Phi}\left( {\mathbf{\sigma}_{n} + \mathrm{\Delta}\mathbf{\sigma}} \right) \approx \underset{z_{n}}{\underbrace{D_{\Phi}\left( \mathbf{\sigma}_{n} \right)}} + J_{n}^{T}\nabla x
	\end{split}
\end{equation}
where $J_{n}$ is a Jacobian matrix, the penalty term can be approximated as
\begin{equation} \label{23}
	\begin{split}
		\left\| {\mathbf{\sigma} - D_{\Phi}\left( {\mathbf{\sigma}_{n} + \mathrm{\Delta}\mathbf{\sigma}} \right)} \right\|^{2} \approx \left\| {\mathbf{\sigma} - z_{n}} \right\|^{2} + \left\| {J_{n}\mathrm{\Delta}\mathbf{\sigma}} \right\|^{2}
	\end{split}
\end{equation}
\begin{figure*}
	\centering
	\subfigure[]{\includegraphics[width=0.95\linewidth]{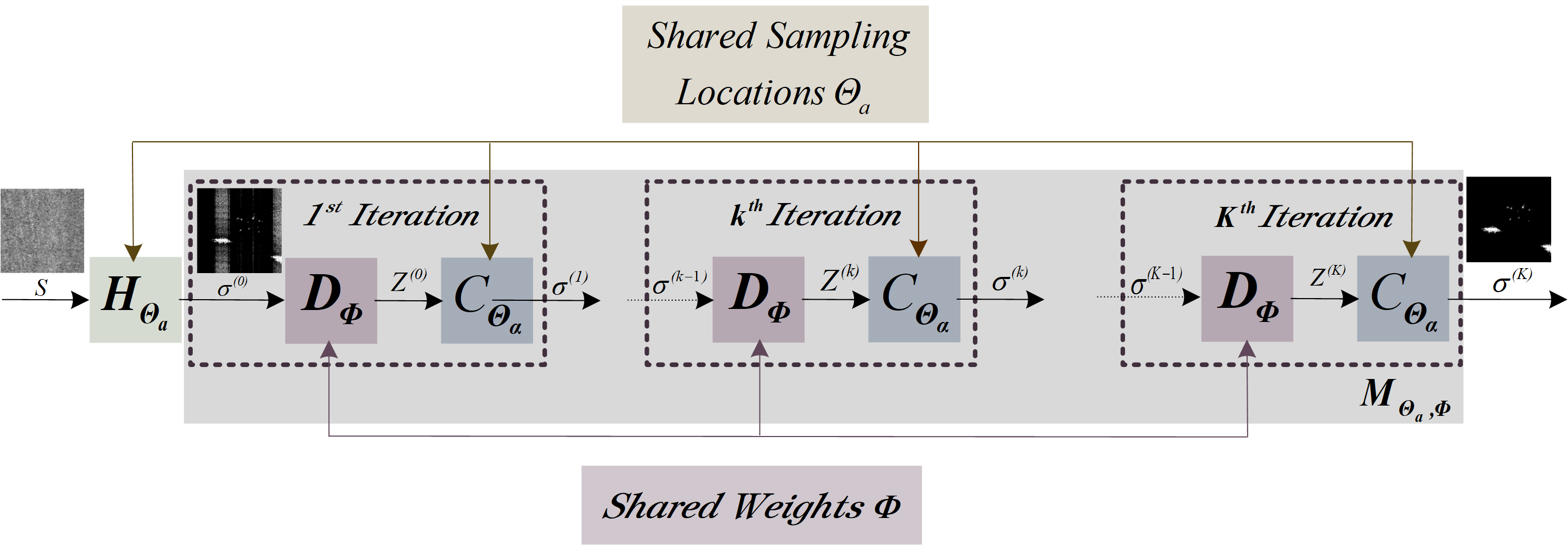} }
	\caption{The unrolled architecture for $K$ iterations, which share weights with the denoising blocks $D_{\Phi}$.}
	\label{fig:Fig4}
\end{figure*}
The above approximation is only valid in the vicinity of $\mathbf{\sigma}_{n}$, hence we obtain the alternating algorithm that approximate (\ref{21}):
\begin{equation} \label{24}
	\begin{split}
		\mathbf{\sigma}_{n + 1} = \left( {H_{\Theta_{a}}H_{\Theta_{a}}^{*} + I} \right)^{- 1}\left( {z_{n} + \mathbf{S} - H_{\Theta_{a}}^{*}\mathbf{\sigma}} \right)
	\end{split}
\end{equation}
\begin{equation} \label{25}
	\begin{split}
		z_{n + 1} = D_{\Phi}\left( \mathbf{\sigma}_{n + 1} \right)
	\end{split}
\end{equation}
here, (\ref{24}) is implemented using a conjugate gradient algorithm:
\begin{equation} \label{26}
	\begin{split}
		\mathbf{\sigma}_{n + 1} = \underset{C_{\Theta_{a}}}{\underbrace{\left( {H_{\Theta_{a}}H_{\Theta_{a}}^{*} + I} \right)^{- 1}}}\left( {H_{\Theta_{a}}\left( \mathbf{S} \right){+ \lambda z}_{n}} \right)
	\end{split}
\end{equation}

This iterative algorithm is unrolled to obtain a deep recursive network $M_{\Theta_{a},\Phi}$, where the weights of the CNN blocks and data consistency blocks are shared across iterations, as shown in Fig.~\ref{fig:Fig3}. Speciﬁcally, the solution to (\ref{21}) is given by
\begin{equation} \label{27}
	\begin{split}
		\hat{\mathbf{\sigma}}\left\{ {\Theta_{a},\Phi} \right\} = M_{\Theta_{a},\Phi}\left( H_{\Theta_{a}}\left( \mathbf{S} \right) \right)
	\end{split}
\end{equation}

Note that once unrolled, the image reconstruction algorithm is essentially a deep network, shown in Fig.~\ref{fig:Fig4}. Thus, the main distinction between MoDL and direct-inversion scheme is the structure of  network $M_{\Theta_{a},\Phi}$.

\subsection{Joint Optimization of Sampling Pattern}\label{sec:3.2}

In this work, we concentrate on optimizing the azimuth sampling pattern denoted by $\Theta_{a}$ in (\ref{11}) and the parameter $\Phi$ of the reconstruction algorithm (\ref{20}) to enhance the quality of the reconstructed images. Drawing inspiration from earlier sampling pattern learning approaches employed in MRI based on compressed sensing algorithms \cite{Sherry2020} \cite{Gozcu2018}, we incorporate these models into the SAR imaging field to optimize the azimuth sampling pattern $\Theta_{a}$, such that
\begin{equation} \label{28}
	\begin{split}
		\left\{ \Theta_{a}^{*} \right\} = {\arg{\min\limits_{\Theta_{a}}{\sum\limits_{i = 1}^{N}\left\| {{\hat{\sigma}}_{i}\left\{ {\Theta_{a},\Phi} \right\} - \sigma_{i}} \right\|_{2}^{2}}}}
	\end{split}
\end{equation}
is minimized. Here ${\hat{\sigma}}_{i}\left\{ {\Theta_{a},\Phi} \right\}$ is the corresponding reconstruction of training image $\sigma_{i};~i = 1,\ldots,N$ in the optimization. 

To improve the reconstruction performance, this paper uses a MF operator model-based deep learning network (MF-JMoDL-Net) framework to jointly optimize both $C_{\Theta_{a}}$ and $D_{\Phi}$ blocks in the MF-JMoDL framework (\ref{21}), which refers to \cite{Aggarwal2020}. Given training data, this framework jointly learns the sampling pattern $\Theta_{a}$ and the CNN parameters $\Phi$ using
\begin{equation} \label{29}
	\begin{split}
		\left\{ {\Theta_{a}^{*},\Phi^{*}} \right\} = {\arg{\min\limits_{\Theta_{a},\Phi}{\sum\limits_{i = 1}^{N}\left\| {M_{\Theta_{a},\Phi}\left( H_{\Theta_{a}}^{*}\left( s_{i} \right) \right) - \sigma_{i}} \right\|_{2}^{2}}}}
	\end{split}
\end{equation}
where $M_{\Theta_{a},\Phi}$ denotes a general deep learning network architecture that includes forward model as well as unrolled architectures denoted by (\ref{27}).

\subsection{MF-JMoDL-Net Architecture}\label{sec:3.3}
Fig.\ref{fig:Fig3} illustrates the framework of the proposed MF-JMoDL, which alternates between CNN blocks $D_{\Phi}$ and data consistency blocks $C_{\Theta_{a}}$ solely dependent on the sampling pattern. We unroll the MF-JMoDL framework for five iterations of alternating minimization to solve (\ref{21}), as depicted in Fig.\ref{fig:Fig4}. The implementation process of the forward operator $H_{\Theta_{a}}^{}$ is shown in Fig.~\ref{fig:Fig2}(b), with the trainable parameters $\Theta_{a} = \left\lbrack {k_{y_{1}}\ldots k_{y_{i}}\ldots k_{y_{N_{a}}}} \right\rbrack^{T}$. Since the operator $\left( {H_{\Theta_{a}}H_{\Theta_{a}}^{} + I} \right)$ is not analytically invertible for complex operators such as SAR echo data, we need to identify a specific gradient algorithm to handle complex forward models.

In contrast to conventional proximal gradient (PG) algorithms \cite{Hammernik2018, mardani2017, putzky2017} which alternate between steepest descent and CNN blocks, this paper chooses the CG algorithm to avoid a high total number of iterations. In comparison, CG blocks provide a faster reduction per iteration by enforcing a more accurate data-consistency constraint at each unfolding step. Without trainable parameters, numerous CG steps can be performed at each unfolding with no memory cost during training. Experiments in \cite{Aggarwal2019} demonstrate that the CG strategy offers improved performance compared to PG and also facilitates the easy incorporation of other image regularization techniques. In this paper, we set the data consistency block $C_{\Theta_{a}}$ to 10 iterations using the CG algorithm.

Regarding the CNN block $D_{\Phi}$, we implement a U-Net model \cite{Ronneberger2015}, which involves four pooling and unpooling layers with 3$\times$3 trainable filters. The parameters of the blocks $D_{\Phi}$ and $C_{\Theta_{a}}$ are optimized to minimize (\ref{21}). We rely on the automatic differentiation capability of TensorFlow to evaluate the gradient of the cost function with respect to $\Theta_{a}$ and $\Phi$.

\subsection{Continuous Optimization Training Strategy}\label{sec:3.4}
In this subsection, we discuss the loss function and backpropagation gradient calculation of the proposed method, which ensure that the optimization training strategy is continuous and differentiable.

\subsubsection{Loss Function}
In this paper, we specify the loss function as the mean square error between ${\hat{\sigma}}_{i}\left\{ {\Theta_{a},\Phi} \right\}$ and the number of iterations as $K$:
\begin{equation} \label{30}
	\begin{split}
		\left\{ {\Theta_{a}^{*},\Phi^{*}} \right\} = {\arg{\min\limits_{\Theta_{a},\Phi}{\sum\limits_{i = 1}^{N}\left\| {M_{\Theta_{a},\Phi}\left( H_{\Theta_{a}}^{*}\left( s_{i} \right) \right) - \sigma_{i}} \right\|_{2}^{2}}}}
	\end{split}
\end{equation}
where $\sigma_{i}$ is the $i^{th}$ target image.

\subsubsection{Backpropagation and Gradient Calculation}
The goal of training is to determine the weight parameter $w$, which is shared across the iterations. The gradient of the cost function with respect to the shared weights can be determined using the chain rule
\begin{equation} \label{31}
	\begin{split}
		{\nabla_{w}\mathcal{L}} = {\sum\limits_{k = 0}^{K - 1}{J_{w}\left( z_{k} \right)^{T}\left( {\nabla_{z_{k}}\mathcal{L}} \right)}}
	\end{split}
\end{equation}
where the Jacobian matrix $J_{w}(z)$ has entries ${\left[ {{J_w}\left( {\rm{z}} \right)} \right]_{i,j}} = \partial {z_i}/\partial {\omega _j}$ and ${z_k}$ is the output of the CNN at the ${k^{th}}$ iteration. We now focus on how to backpropagate through the conjugate gradient (CG) blocks. It is important to note that the CG block does not have any trainable parameters. We have
\begin{equation} \label{32}
	\begin{split}
		{\nabla _{{z_{k - 1}}}}{\cal L} = {J_{{z_{k - 1}}}}{\left( {{\sigma _k}} \right)^T}{\nabla _{{\sigma _k}}}{\cal L}
	\end{split}
\end{equation}
where the Jacobian matrix ${J_z}\left( \sigma  \right)$ has entries ${\left[ {{J_z}\left( \sigma  \right)} \right]_{i,j}} = \partial {\sigma _i}/\partial {z_j}$. The Jacobian matrix is given by
\begin{equation} \label{33}
	\begin{split}
		{J_z}\left( \sigma  \right) = {\left( {{H_{{{\rm{\Theta }}_a}}}H_{{{\rm{\Theta }}_a}}^H + \lambda I} \right)^{ - 1}}
	\end{split}
\end{equation}

Since the Jacobian matrix is symmetric, we have
\begin{equation} \label{34}
	\begin{split}
		{{\nabla _{{z_{k - 1}}}}{\cal L}} = {\left( {{H_{{{\rm{\Theta }}_a}}}H_{{{\rm{\Theta }}_a}}^* + \lambda I} \right)^{ - 1}}\left( {{\nabla _{{\sigma _k}}}{\cal L}} \right)
	\end{split}
\end{equation}

We can evaluate the above expression using a CG algorithm, running until convergence, determined by the saturation of the data consistency cost. It should be noted that the above gradient calculation is only valid if we implement ${\left( {{H_{{{\rm{\Theta }}a}}}H{{{\rm{\Theta }}_a}}^* + \lambda I} \right)^{ - 1}}$ or, equivalently, let the CG algorithm converge. This result demonstrates that the gradients can be backpropagated through the CG block. We rely on the variable sharing strategies in TensorFlow to implement the unrolled architecture. The parameters of the networks at each iteration are initialized and updated together.

\section{Experiments and Discussions}\label{sec:4}
\begin{figure}[t]
	\vspace{-1.0em}
	\centering
	\subfigure[] {
		\includegraphics[width=0.45\linewidth]{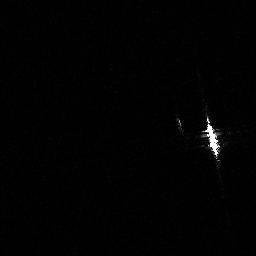}}
	\subfigure[]{
		\includegraphics[width=0.45\linewidth]{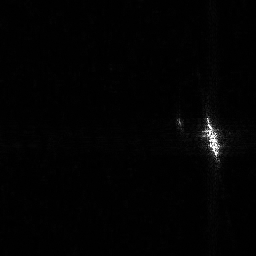}}\\
	\caption{(a) represents oiriginal SAR amplitude image. (b) represents reconstruction SAR amplitude image.}
	\vspace{-0.5em}
	\label{fig:Fig5}
\end{figure}
\subsection{Experiment Settings}
\subsubsection{Dataset}
In the following experiments, the training and testing data are derived from SSDD sea surface ships SAR images. We use a total of 1300 SAR patches, with sizes standardized to 256$\times$256 pixels. We categorize the images into three groups based on scene sparsity: off-shore ships, coastal ships, and strait. The distribution of training and testing datasets for each category is shown in Table \ref{tab:tab1}.

\begin{table}[t]
	\centering
	\caption{Configuration of Training and Testing Dataset}
	\label{tab:tab1}
	\begin{tabular}{ccc}
		\toprule
		\textbf{}  & \textbf{Training Numbers} & \textbf{Testing Numbers}\\ \toprule
		\textbf{Off-shore Ships}         			& 600            & 100\\
		\textbf{Coastal Ships}             			& 200   	      & 50\\
		\textbf{Strait}          & 200    	      & 50\\
		\bottomrule		
	\end{tabular}
\end{table}
Similar to \cite{Zhao2021}, we generate unlabeled training samples by downsampling the original echo data, adding system noise, and introducing echo phase disturbances. This method not only reduces the amount of data and increases imaging efficiency, but also improves the robustness and reliability of the algorithm. The parameters used in these processes are provided in Table \ref{tab:tab2}.

Based on the parameters in Table \ref{tab:tab2}, the amplitude map of the SSDD dataset is used to generate echoes as follows: let the amplitude image of the SSDD dataset be $A$, the image after adding a random phase is ${A_{SLC}}$, and the point target echo ${S_p}$ generated in Table \ref{tab:tab2}. Then, the original echo of the $i$-th image ${S_i} = {S_p}{\rm{*}}{A_{SLC}}$. ${S_i}$ is compared with the original image after imaging as shown in Fig.~\ref{fig:Fig5}.
\begin{table}[t]
	\centering
	\caption{MAIN PARAMETERS OF SAR SYSTEMS}
	\label{tab:tab2}
	\begin{tabular}{cc}
		\toprule
		\textbf{Parameter}          	 & \textbf{Value} \\ \midrule
		Range bandwidth        		     & 150 MHz        \\
		Range sampling rate              & 200 MHz        \\
		Pluse duration              	 & 1 $\mu$s	  \\
		Range chirp rate           		 & 200e12 Hz/s    \\
		Doppler rate           			 & 256.1772 Hz/s  \\
		Pulse repetition frequency       & 200 Hz         \\
		Carrier frequency            	 & 9.6 GHz        \\
		Wavelength                       & 0.0375 m       \\
		Azimuth antenna length           & 2 m            \\ 
		Platform height				     & 10000 m		  \\
		Equivalent velocity				 & 200 m/s		  \\ \bottomrule
	\end{tabular}
\end{table}

\subsubsection{Evaluation Index Definitions}
Structural similarity (SSIM) and peak signal-to-noise ratio (PSNR) are used to quantitatively evaluate the experiment results. Let $\mathbf{\sigma} $ be the result of target imaging and $\mathbf{\hat \sigma} $ be the target label image. The definitions of these evaluation indexes are as follows:
\begin{equation} \label{35}
	\begin{split}
		SSIM\left( {\hat \sigma ,\sigma } \right) = \frac{{\left( {2{\mu _{\hat \sigma }}{\mu _\sigma } + {C_1}} \right)\left( {2{\rho _{\hat \sigma }}{\rho _\sigma } + {C_2}} \right)}}{{\left( {\mu _{\hat \sigma }^2\mu _\sigma ^2 + {C_1}} \right)\left( {\rho _{\hat \sigma }^2 + \rho _\sigma ^2 + {C_2}} \right)}}
	\end{split}
\end{equation}
where ${\mu _{\hat \sigma }}, {\mu _\sigma }, {\rho _{\hat \sigma }}, {\rho _\sigma }$,  and ${\rho _{\hat \sigma }}{\rho _\sigma }$ are the local means, standard deviations,  cross covariance for images $\mathbf{\hat \sigma} $ and $\mathbf{\sigma}$, respectively. ${C_1} = {\left( {{K_1}L} \right)^2}$, $L$ is the variation range of the pixel value, as for 8-bit grayscale image, $L = 255$, ${K_1} \ll 1$, ${C_2} = {\left( {{K_2}L} \right)^2}$, ow we set $K_1=0.01$ and $K_2=0.03$
\begin{equation} \label{36}
	\begin{split}
		PSNR = 10{\log _{10}}\frac{{{\rm{max}}\sigma _2^2}}{{\hat \sigma  - \sigma _2^2}}
	\end{split}
\end{equation}

\subsection{Results and Analysis}

To demonstrate the performance of image reconstruction and enhancement provided by MF-JMoDL-Net, experiments are conducted under various conditions, such as different sparsity scenes, different sampling rates, and different sampling patterns. Additionally, the azimuth ambiguity suppression capability of the proposed method is compared with the local azimuth ambiguity-to-signal ratio estimation (LAASR) \cite{long_azimuth_2020} ambiguity suppression proposed by Long. The design of the azimuth sampling pattern using the proposed method is compared with staggered \cite{Villano2012, Villano2014a, Villano2014b, Villano2015, Villano2017}, uniform sampling, and Poisson-disk sampling methods \cite{ Yang2019}.
\begin{figure}[t]
	\centering
	\includegraphics[width=0.95\linewidth]{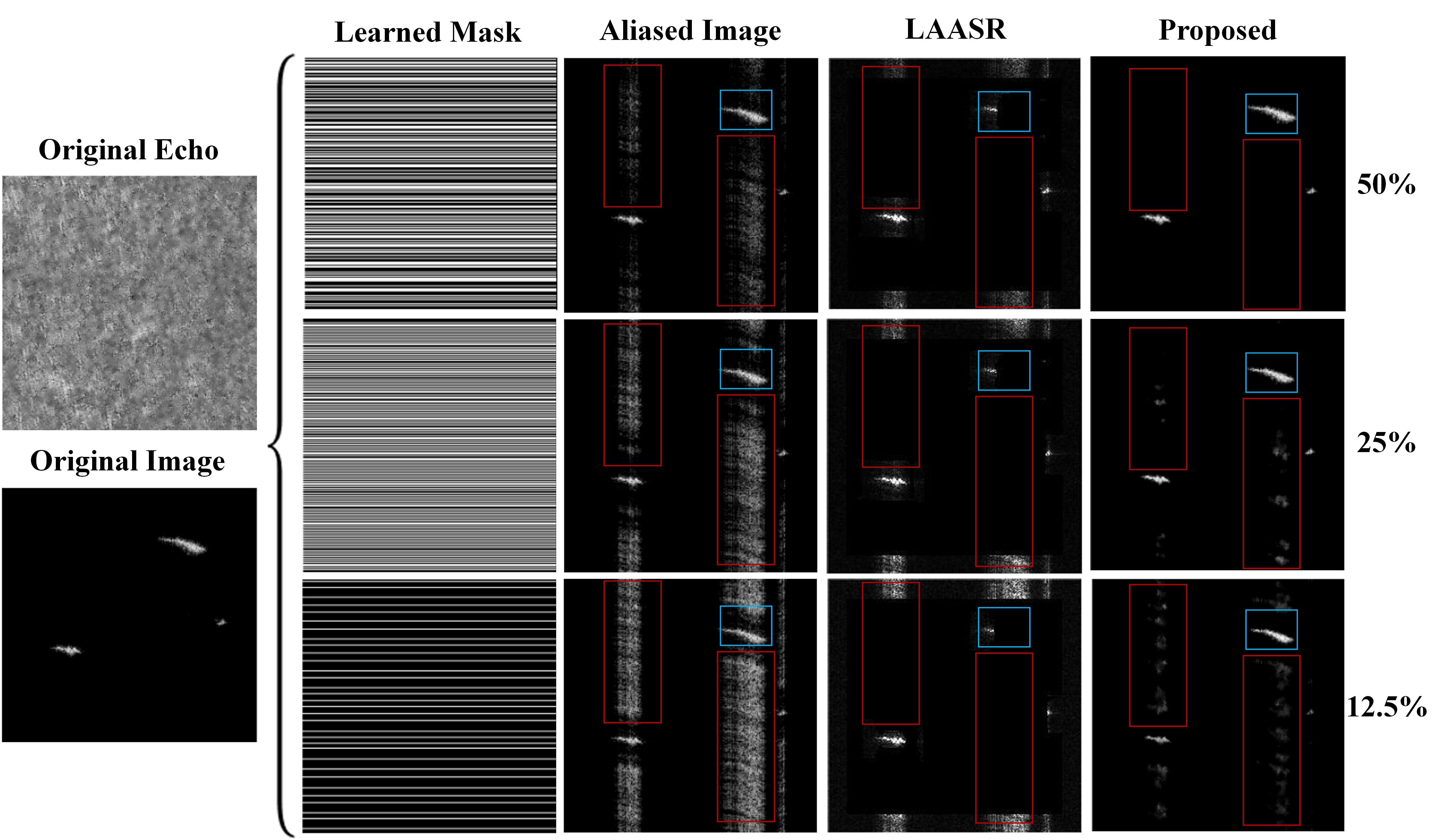}
	\caption{The performance in terms of azimuth ambiguity suppression under 50$\%$, 25$\%$ and 12.5$\%$ azimuth undersampling rates.}
	\vspace{-2mm}
	\label{fig:Fig6}
\end{figure}

\subsubsection{Results Under Different Azimuth Undersampling Rates}
\newcommand{\tabincell}[2]{\begin{tabular}{@{}#1@{}}#2\end{tabular}}
\begin{table}[htbp]
	\centering
	\caption{results of different undersampling rates and scenes}
	\label{tab:tab3}
	\begin{tabular}{cccccc}
		\toprule
		\textbf{\tabincell{c}{sampling\\ rate($\%$)}}     & \textbf{Scence} & \textbf{SSIM($\%$)} & \textbf{\tabincell{c}{Undersamping\\ PSNR (dB)}} & \textbf{\tabincell{c}{Reconstruction\\ PSNR(dB)}} &\textbf{\tabincell{c}{Increase\\ PSNR (dB)}} \\ \midrule
		50      &Off-shore ships  	&0.932306	&25.51  &37.50  &11.99   \\
		25      &Off-shore ships      &0.845824   &21.50  &33.36  &11.86   \\
		12.5    &Off-shore ships      &0.777514   &18.25	&30.27  &12.02   \\ 
		50		&Coastal ships          &0.724638   &14.01	&18.69	&4.68    \\
		50		&Strait 					&0.699019   &13.63  &18.04  &4.41    \\ \bottomrule
	\end{tabular}
\end{table}
We use data with 50\%, 25\%, and 12.5\% azimuth undersampling rates to test the influence of the undersampling rate on both the LAASR method and the proposed method. The learned mask, which represents sampling locations, is fixed with the same pattern obtained by the proposed MF-JMoDL-Net. The performance in terms of azimuth ambiguity suppression is shown in Fig.~\ref{fig:Fig6}.

In Fig.~\ref{fig:Fig6}, the red boxes indicate locations with severe ambiguity, while the blue boxes show the original target structure. In the case of a 50\% undersampling rate, the method proposed in this paper demonstrates better effectiveness in azimuth ambiguity suppression. It is evident that our method's performance deteriorates as the sampling rate decreases. However, the LAASR method is not affected by undersampling rates and eliminates true targets. The LAASR method relies on the AAP, which fixes the locations of the ambiguity region and removes them awkwardly. In other words, the LAASR method may misjudge targets as ambiguity and eliminate all or part of them. In contrast, the proposed method preserves the target structures more intelligently.
\begin{table}[htbp]
	\centering
	\caption{results of different undersampling pattern}
	\label{tab:tab4}
	\begin{tabular}{ccccc}
		\toprule
		\textbf{\tabincell{c}{sampling\\ rate($\%$)}}     & \textbf{SSIM($\%$)}   & \textbf{\tabincell{c}{Undersamping\\ PSNR (dB)}} & \textbf{\tabincell{c}{Reconstruction\\ PSNR(dB)}} & \textbf{\tabincell{c}{Increase\\ PSNR(dB)}}  \\ \midrule
		Proposed      &0.932306      &25.51   &37.50  &11.99  \\
		Poisson-disk  &0.760917      &20.67   &26.04  &5.28   \\
		Staggered     &0.747645      &21.63   &25.06  &3.43   \\ \bottomrule
		
	\end{tabular}
\end{table}
\begin{figure}[htbp]
	\centering
	\subfigure[Results of off-shore ships under 50$\%$ undersampling rate.]{
		\includegraphics[width=0.8\linewidth]{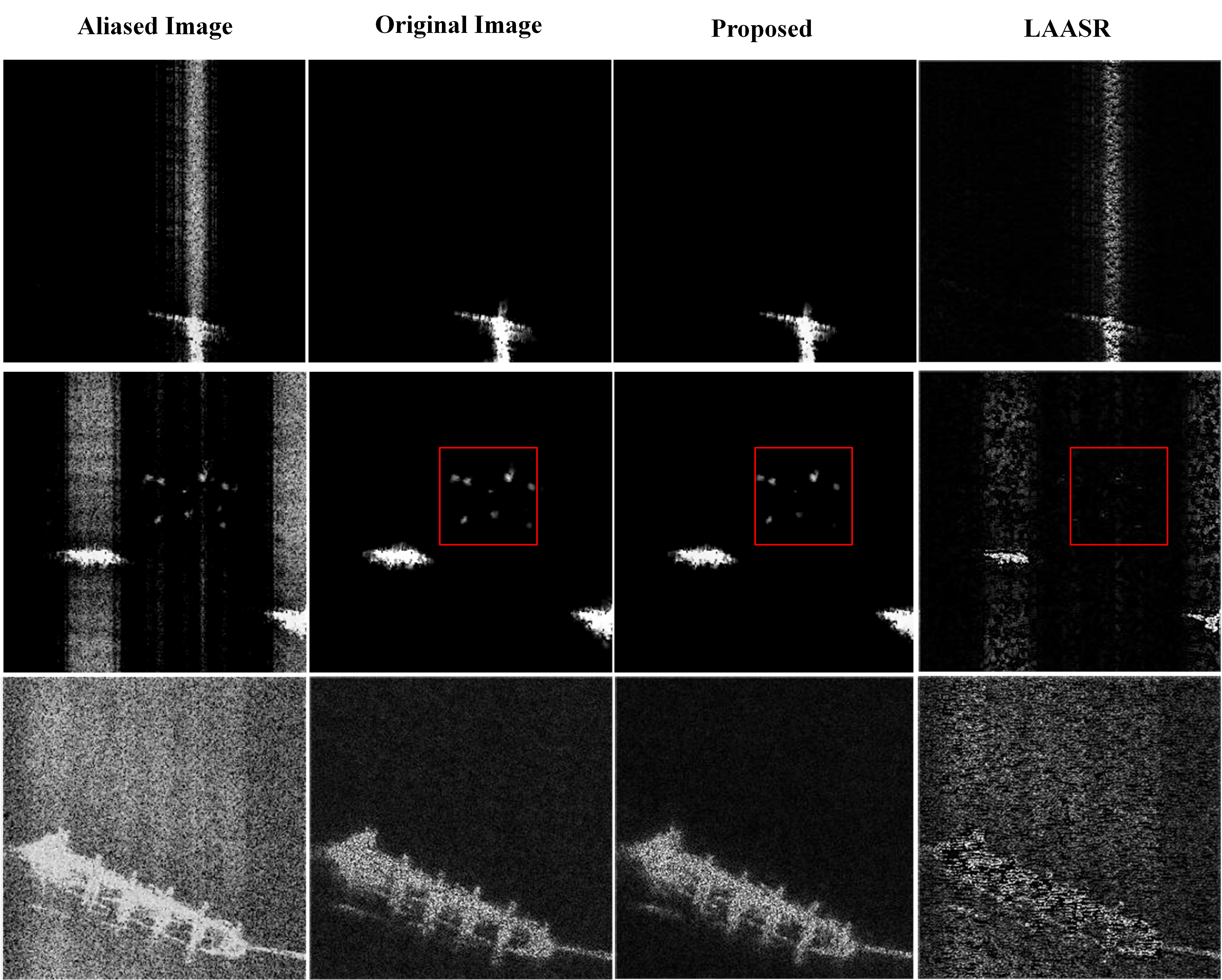}
	}\label{fig:Fig7(a)}
	\subfigure[Results of coastal ships under 50$\%$ undersampling rate.]{
		\includegraphics[width=0.8\linewidth]{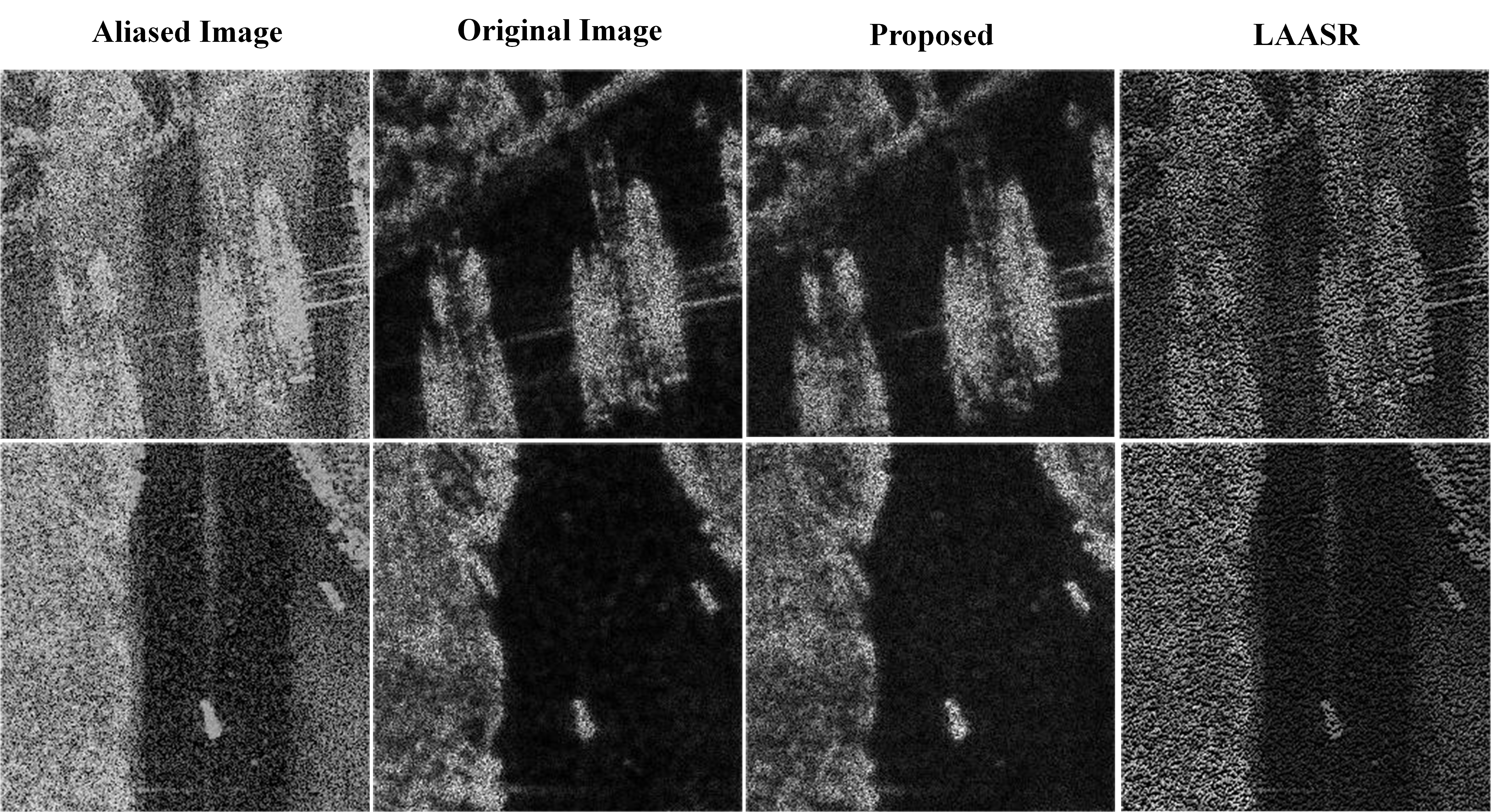}
	}\label{fig:Fig7(b)}
	\subfigure[Results of strait under 50$\%$ undersampling rate.]{
		\includegraphics[width=0.8\linewidth]{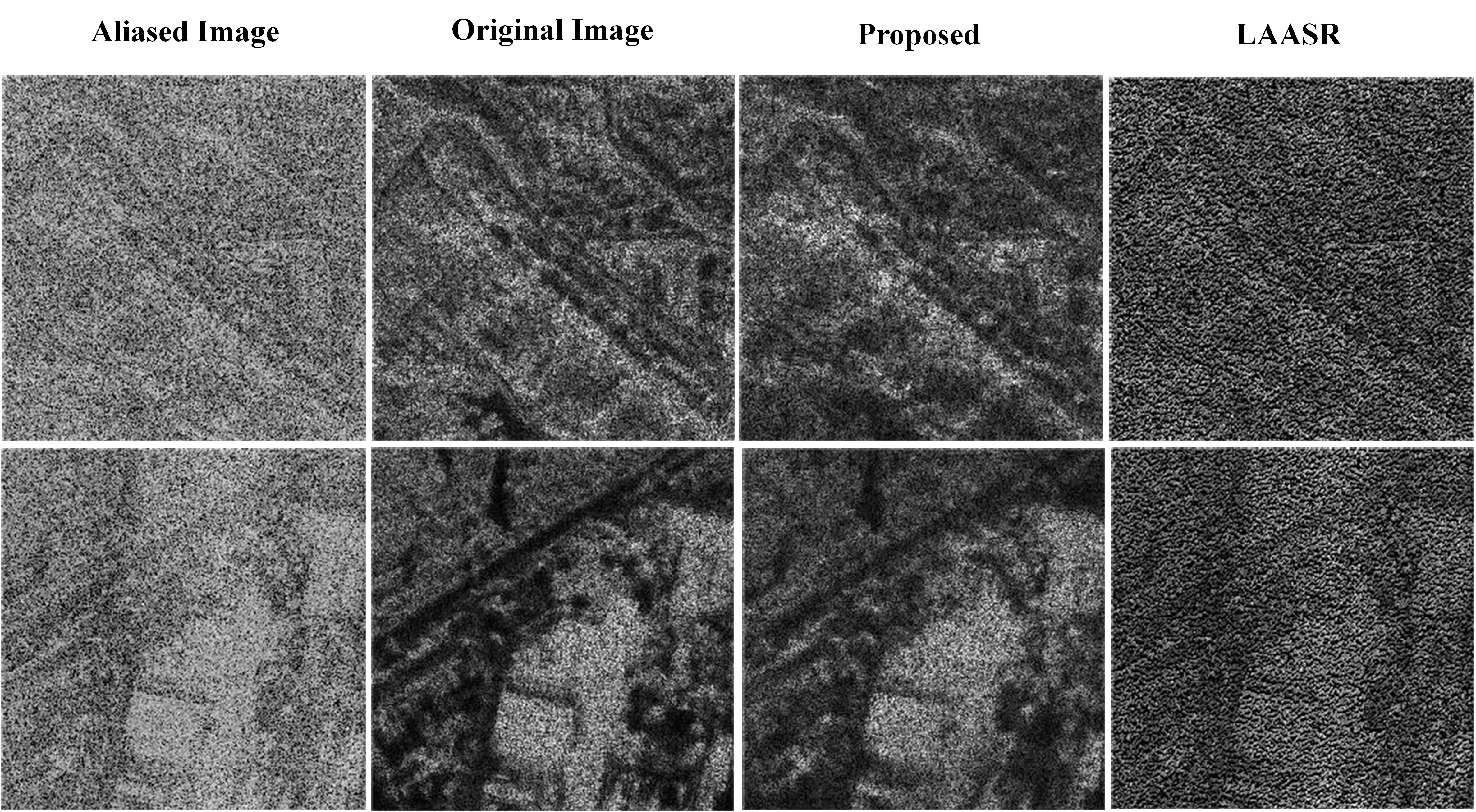}
	}\label{fig:Fig7(c)}
	\caption{The azimuth ambiguity suppression and image reconstruction results of proposed method and LAASR under 50$\%$ sampling in different scenes.}
	\label{fig:Fig7}
\end{figure}

\subsubsection{Results Under Different Scenes} 

Fig.\ref{fig:Fig7} presents the azimuth ambiguity suppression and image reconstruction results of the proposed method and LAASR under a 50\% sampling rate for off-shore ships scenes in Fig.\ref{fig:Fig7}(a), coastal ships scenes in Fig.\ref{fig:Fig7}(b), and strait scenes in Fig.\ref{fig:Fig7}(c), respectively. As the scene sparsity decreases, the ambiguity suppression performance of the proposed method remains stable. However, the LAASR method not only fails to suppress ambiguity effectively, but also significantly deteriorates the structural similarity of the original images. 

To visually display the performance trends of the proposed method, the curves of SSIM, undersampling PSNR, reconstruction PSNR, and increased PSNR values under different undersampling rates are presented in Table \ref{tab:tab3}. From the experimental results in Table \ref{tab:tab3}, the PSNR remains stable as the sampling rate decreases, even though SSIM declines with the sampling rate.

\begin{figure}[t]
	\centering
	\includegraphics[width=0.95\linewidth]{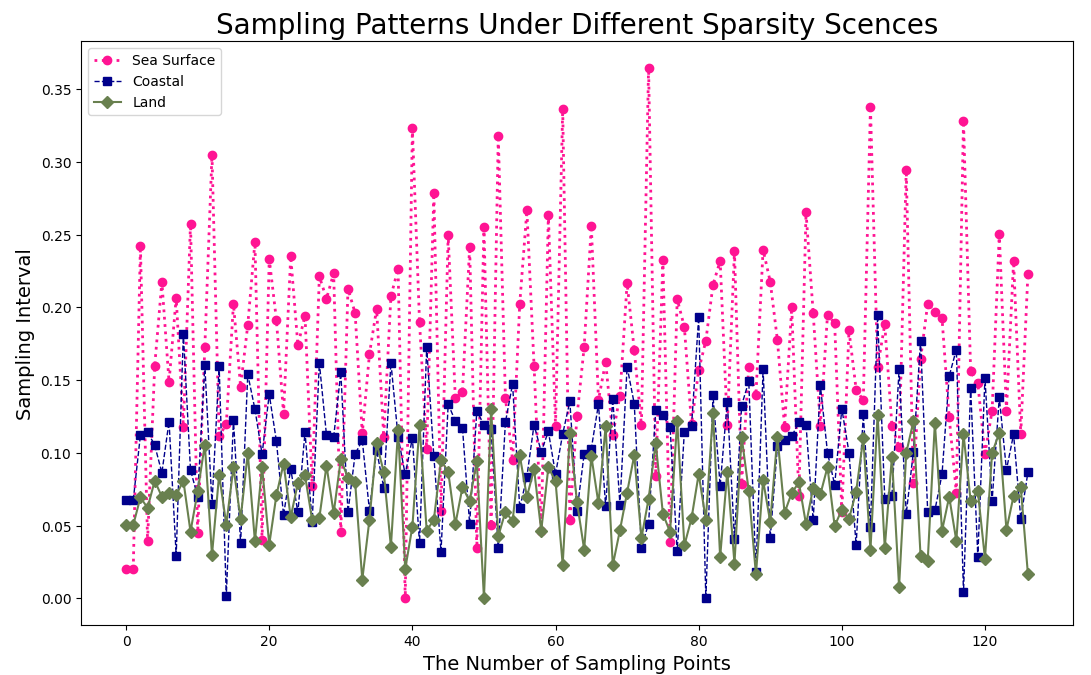}
	\caption{Trained sampling intervals under different sparsity scenes.}
	\vspace{-2mm}
	\label{fig:Fig8}
\end{figure}

Fig.\ref{fig:Fig8} displays the distribution of distances between sampling positions trained by the proposed MF-JMoDL-Net under typical representative scenes with three degrees of sparsity. In Fig.\ref{fig:Fig8}, pink dots represent the off-shore ship scenes, blue squares represent coastal scenes, and green diamonds represent strait scenes. Clearly, the intervals between sampling positions change gradually as the scene sparsity decreases. Consequently, it is also worth discussing and verifying whether the optimal sampling pattern approaches uniform sampling when the scene sparsity reaches its lowest level.

\subsubsection{Results Under Different Sampling Pattern}

We utilize the off-shore ship dataset in Table \ref{tab:tab1} as the unified training target, and compare the reconstruction results of the proposed method with the staggered sampling pattern and Poisson-disk sampling pattern on the azimuth ambiguity image, as shown in Table \ref{tab:tab4}.

\begin{figure}[t]
	\centering
	\includegraphics[width=0.95\linewidth]{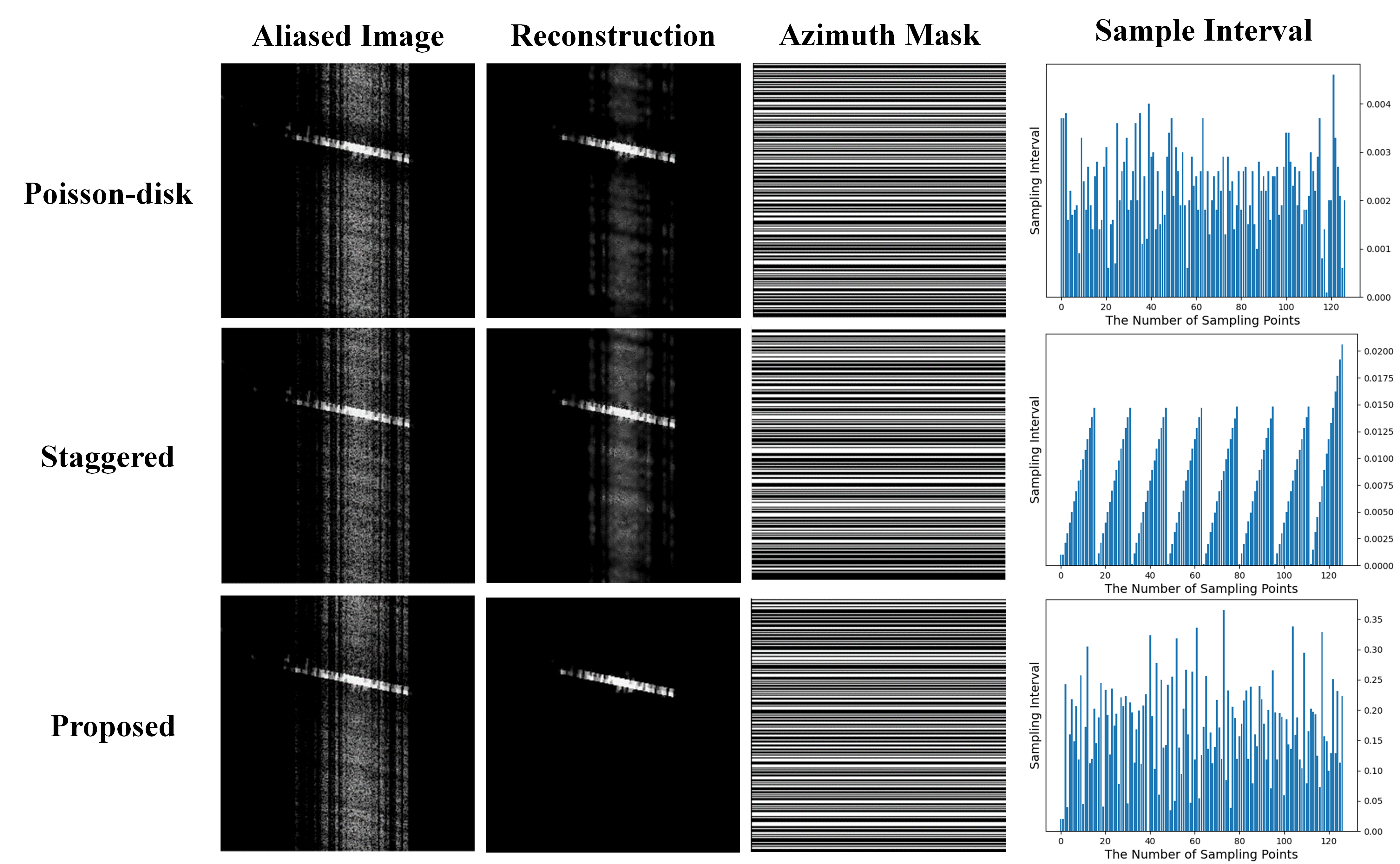}
	\caption{Results of different undersampling patterns.}
	\vspace{-2mm}
	\label{fig:Fig9}
\end{figure}

From the results in Table \ref{tab:tab4} and Fig.~\ref{fig:Fig9}, the sample interval is shown in the last row. Clearly, the method proposed in this paper performs best in terms of SSIM and PSNR improvement.

\section{Conclusion}\label{sec:5}
In this paper, we present a joint optimization framework for sparse strip SAR imaging algorithms and azimuth undersampling patterns, namely MF-JMoDL-Net. The architecture of the MF-JMoDL-Net is constructed by introducing the MF approximate measurement operator and inverse MF operator into the JMoDL-based deep network, which serves as the solution for the proposed MF-JMoDL-based imaging model. Through end-to-end training, the extraction and utilization of echo data information are realized, enabling the acquisition of optimal azimuth sampling patterns and the suppression of corresponding ambiguity caused by azimuth undersampling. To verify the proposed method, we conducted experiments and comparisons under various conditions, including different sampling rates (50\%, 25\%, and 12.5\%), different sparsity levels of scenes, and sampling patterns (staggered and Poisson-disk). All of these experiments and comparisons demonstrate the effectiveness and superiority of the proposed MF-JMoDL-Net compared to existing methods.

\bibliography{mybibfile}

\end{document}